\documentclass{pic2012}
\def\runauthor{Lundberg, Olof}
\def\shorttitle{Calibration Systems of the ATLAS Tile Calorimeter}
\begin{document}

\title{CALIBRATION SYSTEMS OF THE ATLAS TILE CALORIMETER
}

\author{Olof Lundberg}

\address{Department of Physics, Stockholm University, 106 91 Stockholm, Sweden\\
E-mail: olof.lundberg@fysik.su.se\\
(on behalf of the ATLAS Collaboration)}

\maketitle

\abstracts{TileCal is the hadronic calorimeter covering the most central region of the ATLAS experiment at the LHC. This sampling calorimeter uses iron plates as absorber and plastic scintillating tiles as the active material. A multi-faceted calibration system allows to monitor and equalize the calorimeter response at each stage of the signal production, from scintillation light to digitization. This calibration system is based on signal generation from different sources: a Cs radioactive source, laser light, charge injection and minimum bias events produced in proton-proton collisions. A brief description of the different TileCal calibration systems is given and the latest results on their performance in terms of calibration factors, linearity and stability are presented.
}

\section{Introduction} 
ATLAS [1] is a general purpose experiment installed at the Large Hadron Collider
(LHC) at CERN. The Tile Calorimeter (TileCal) [2] is the hadronic calorimeter
covering the most central region, $|\eta| < 1.7$ of the ATLAS detector. The focus of
TileCal is to perform precise measurements of hadrons, jets, taus and the missing
transverse energy as well as to provide input signal to the Level 1 Calorimeter
Trigger. TileCal is a sampling device which uses iron plates as absorber and plastic 
scintillating tiles as the active material. The scintillator light is transmitted by
wavelength shifting fibers and read out by photomultiplier tubes (PMTÕs). Cells
are defined by grouping these fibers into bundles and coupling them to a given
PMT. These PMTÕs can be read out in either high gain or low gain amplification
depending on the signal strength. Each PMT also has a separate integrator system which integrates the PMT current over time~\cite{Integrator}. The calibration systems described in the following sections are designed to test all the steps of the readout chain: The optical properties of the scintillators, the PMT gains and the front-end electronics. 

\section{Monitoring and calibrations of the TileCal}
TileCal calibration relies on several different dedicated systems: (i) calibration of
the tile optic components using movable cesium radioactive gamma sources; (ii)
calibration of the photomultiplier tube (PMT) gains and linearities using a laser
calibration system; (iii) calibration of the front-end electronic gains using a charge injection system (CIS); (iv) monitoring of the channels using minimum bias events with the integrator system.\\
These systems in combination makes it possible to monitor and equalize the
response of the calorimeter at each stage of the signal generation, from the scintillators, to the PMTs, to the electronics.

\subsection{The cesium system}
In dedicated Cesium runs, ${}^{137}$Cs sources with activities of around 330 MBq, emitting 662 keV $\gamma$-rays are hydraulically driven through a system of steel tubes traversing the cells. The integrator circuits of each channel read out the cell when it is
traversed by the source. The total integrated current read out is normalised to
the cell size. The cesium system calibrates the entire readout chain, and provides an absolute scale determination. 
These scans can be used to diagnose the optical instrumentation, to
measure the response of each cell and to equalize the response of the calorimeter at
the electromagnetic scale. The Cesium calibration allows to test the optical chain
with precision better than 0.3\%.

\subsection{The laser system}
The laser calibration system is dedicated to monitoring and calibrating the gain and
linearity of each Tile PMT. It is also useful in timing studies. The laser
provides a beam of 532 nm light emitted in short ($\sim$ 15 ns) pulses simulating physics
signals, with power enough to saturate all Tile readout channels. A dedicated set
of optical elements propagates the original light beam to every photomultiplier.
This is performed in special dedicated laser runs done several times each week.
The laser system is also used in empty orbits during physics runs, to monitor the
short-term PMT gain variation and to monitor timing. The typical precision of the
laser system is better than 0.5\% on the gain variation [4].

\subsection{The charge injection system}
The CIS is a resident part of each front-end electronics channel and is used to
measure the pC/ADC conversion factor for the digital readout of physics 
and laser calibration data. It simulates physics signals in the different Tile-
Cal channels by injecting a known charge and measuring the electronic response.
CIS calibration constants are calculated from dedicated charge injection runs, and
are used for monitoring purposes and for correct energy reconstruction. The CIS
constants are stable over time: in the period February-June 2011 the mean value
of these constants varied by 0.5\%. The precision of the CIS calibrations is within
0.7\%.

\subsection{The integrator system}
In the high energy proton-proton collisions at the LHC the dominating interactions
are soft parton interaction, or Minimum Bias (MB) events.
The integrator system of each PMT integrates the PMT gain over time and is used to measure the signal of the MB interactions during proton-proton
collisions. The MB interactions are proportional to the instantaneous luminosity
and the measured integrated current is used monitor the channel stability. It can
also be used to measure luminosity.

\section{Combined calibrations}
When combining results from all calibration systems one can determine what causes
changes in the overall detector response. In the first two years of operation an
updrift in the PMT gains was observed. This effect disappeared in 2011 when the
LHC started operating at higher luminosity than previously. Since then a downdrift is observed during the time the beam is on, and a slow recovery in the time
the beam is off. This effect is seen by the Cesium scans, the integrator system and
the laser calibrations. That all three systems show the same behavior leads to the
conclusion that it is caused by drift of the PMT gains. It is mostly affecting PMTÕs
at lower radius which are also the ones which receive the most scintillator light.

\begin{figure}[h!]
  \centering
     \includegraphics[width=0.58\textwidth]{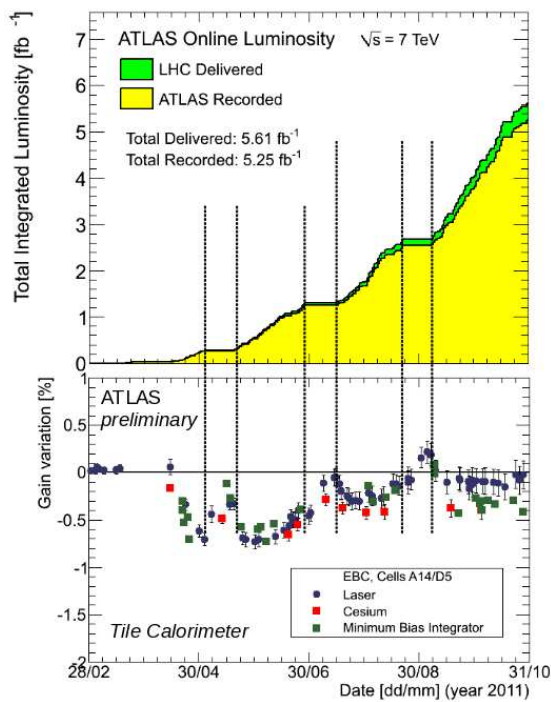}
  \caption{The evolution of ATLAS total integrated luminosity (top) and the evolution of the response of a cell at low radius. The response is measured separately by the Cesium, Laser and Minimum Bias calibration systems.}
\end{figure}

\section{Electronic noise}
Electronic noise is measured in dedicated pedestal runs, where all channels are
read out in periods with no collisions. The noise is mainly used for monitoring
detector performance, for setting trigger thresholds and as input to jet reconstruction algorithms.
In the winter 2011-2012 fourty out of the 256 TileCal modules had their Low Voltage
Power Supply (LVPS) exchanged for a newer type. The main aim was to reduce the
number of trips, and also to reduce noise levels. Cells in modules with newer LVPS
type have on average around 13\% lower noise compared to the previous LVPS,
which is demonstrated in Fig. 2. The number of trips in 2012 is (as of september
2012) over 4000, whereof only 4 occurred in modules with new LVPS.

\begin{figure}[h!]
  \centering
     \includegraphics[width=0.72\textwidth]{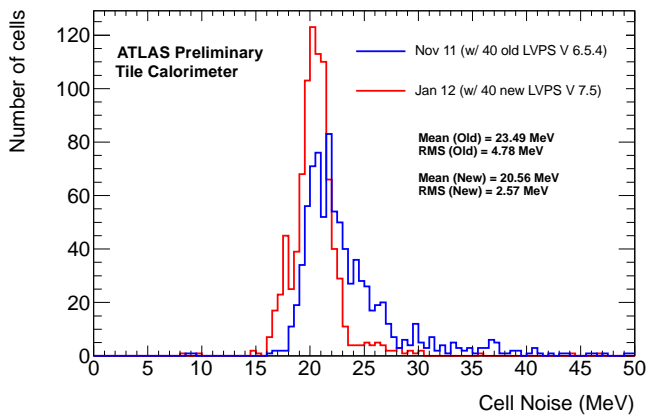}
  \caption{RMS of the electronic noise for all cells in the fourty modules which had their LVPS changed in winter 2011-2012, comparing a pedestal run from November 2011 to a run from January 2012.}
\end{figure}

\section{Conclusions}
The calibration systems of the ATLAS TileCal have been presented. The CIS, laser
and cesium calibration systems allow for calibrations and monitoring of calorimeter
response with a 0.5-1.0\% precision. Analysis of the combined calibrations can be
used to gain detailed insight in what causes variations in detector response. The
stability and linearity of the calibration constants have been presented. A study of
the lowering of the electronic noise using the new LVPS has also been introduced.


\begin{thebibliography}{0}

\bibitem{atlas}  
The Atlas Collaboration (2008); The ATLAS experiment at the CERN Large Hadron Collider, JINST {\bf 3}, S08003

\bibitem{ready}
The ATLAS Collaboration (2010); Readiness of the ATLAS Tile Calorimeter for LHC collisions, Eur. Phys. J. C {\bf70}

\bibitem{Integrator}
G. Gonzalez Parra (2011); Integrator based readout in Tile Calorimeter of the ATLAS experiment, ATL-TILECAL-PROC-2011-010, CERN 

\bibitem{Laser} 
V. Giangobbe (2011); The TileCal Laser calibration system, ATL-TILECAL-PROC-2011-007, CERN


\end{thebibliography}
\end{document}